\documentclass[3p,number]{elsarticle}

\usepackage{hyperref,booktabs,amsmath,tablefootnote,verbatim}
\usepackage[flushleft]{threeparttable}

\journal{Pervasive and Mobile Computing}

\begin{document}

\begin{frontmatter}

\title{Multidevice Mobile Sessions: A First Look}

\author[aalto]{B. Finley\corref{cor1}}
\ead{benjamin.finley@aalto.fi}

\author[aalto]{T. Soikkeli}
\cortext[cor1]{Corresponding author}
\ead{tapio.soikkeli@aalto.fi}

\address[aalto]{Department of Communications and Networking, Aalto University, Otakaari 5, Espoo, Finland}

\begin{abstract}
The increasing number of users with multiple mobile devices underscores the importance of understanding how users interact, often simultaneously, with these multiple devices. However, most device based monitoring studies have focused only on a single device type. In contrast, we study the multidevice usage of a US-based panel through device based monitoring on panelist's smartphone and tablet devices. We present a broad range of results from characterizing individual multidevice sessions to estimating device usage substitution. For example, we find that for panelists, 50\% of all device interaction time can be considered multidevice usage.
\end{abstract}

\begin{keyword}
Mobile Device\sep Multidevice Usage\sep Mobile Session\sep Device Substitution
\end{keyword}

\end{frontmatter}


\section{Introduction}
Today a growing number of users own multiple mobile devices including, for example, a smartphone, tablet, and smartwatch. This proliferation of devices can cause significant changes in usage behavior both in terms of individual sessions and general usage. For example in general, users can shift certain activities from an existing device (i.e. smartphone) to a new device (i.e. tablet), or the new device can prompt totally novel usage. While inside single sessions, users can use multiple devices either in combination or simply simultaneously, and these complex sessions can be described as so called multidevice usage sessions.

Understanding these types of changes is increasingly important for a variety of stakeholders in the mobile ecosystem. For example, mobile application developers are interested in improving the quality of activities that start on one device and continue on another device. A concrete implementation is already seen in web browsers that can sync open web browser tabs across devices\footnote{https://support.apple.com/en-us/HT202530}. Whereas mobile advertisers are interested in understanding how users divide their time between mobile device types to help guide mobile advertising strategies. For example, advertisers can understand the relative ad inventories (number of available impressions) between specific device types of users by considering usage times and typical ad refresh rates.

However, previous studies on mobile device usage have not thoroughly examined this multidevice emergence for a variety of reasons. Many mobile device usage studies primarily tracked only a single device type (typically a smartphone) and often only captured a single mobile platform (typically Android) \cite{bohmer2011,falaki2010,oliver2010,soikkeli2013}. Several studies have captured multidevice usage but have used interview, survey, or diary collection methods \cite{dearman2008,google2012,microsoft2013,jokela2015,muller2015} that are susceptible to recall bias \cite{dereuver2012}. Other studies have captured multidevice usage but only from a single service provider view point such as multidevice search \cite{montanez2014, wang2013}. Furthermore few studies have explicitly investigated device type substitution (shifting usage to a new device or novel usage of new device) and again these studies typically rely on survey or diary methods \cite{shmorgun2013}.

In contrast, we examine multidevice (smartphone and tablet) usage of a US based user panel from February 2015. Our data collection is device based (panelists installed custom device based monitoring applications) and multi-platform (Android and iOS). Furthermore, we study device type substitution by utilizing a supplementary (smartphone only) user panel also from February 2015. To our knowledge, this is the first study of a multidevice user panel with device based monitoring.

We make the following concrete contributions in this work.
\begin{itemize}
  \item We present a simple method for grouping app sessions into multidevice usage sessions based on Allen's interval algebra.
  \item We describe the statistical properties of these multidevice sessions on both an overall and per user level. 
  \item We present and utilize a novel method for studying the diversity of temporal patterns in multidevice sessions through a matrix representation.
  \item We examine if users use their smartphones differently during multidevice sessions.
  \item We examine the usage substitution effects between smartphones and tablets by utilizing a complementary panel of users without a tablet device.
\end{itemize}
The remaining sections of the paper are organized as follows. Section \ref{background} discusses related work and introduces Allen's interval relations, while Section \ref{datasets} introduces the panel datasets. Section \ref{usage_session} details the different session definitions and the aggregation of individual app sessions into usage sessions. Section \ref{results} describes and discusses the primary results. Section \ref{discussion} discusses some general implications and the replicability and generalizability of the results and finally Section \ref{conclusions} presents the conclusions and future work.

\section{Background} \label{background}
\subsection{Related Work} \label{sect_rel_work}
Related work can be primarily divided into single device usage studies that use a similar device based monitoring method, multidevice studies that also characterize multidevice usage, and multidevice interaction enablers and frameworks.

\subsubsection{Single Device Usage Studies with Device Based Monitoring}
Eagle and Pentland \cite{eagle2006} pioneered device based monitoring with mobile phones in their Reality Mining study, though they did not explicitly focus on studying mobile device usage. B\"{o}hmer et al. \cite{bohmer2011} analyzed mobile application usage in one of the first large scale mobile usage studies. They used a similar concept of a mobile app session described by an app moving to the foreground or background of the device. Furthermore, they extended the concept to describe an app chain (sequential app sessions with a maximum timeout value between sessions of 30s). These app chains are similar to our concept of single device usage sessions. Soikkeli et al. \cite{soikkeli2013} had a similar approach to \cite{bohmer2011} with 1s and 30s maximum timeout values for app chains, or usage sessions in their terminology. Falaki et al. \cite{falaki2010} also based their analysis on foreground application sessions with an extension to interaction sessions. However, an interaction session was defined as continuous screen-on time period (essentially a timeout value of 0s). Oliver \cite{oliver2010} also studied this type of interaction session. 

All the mentioned studies included only smartphone usage, in some cases because tablets were not yet widespread. Also, the data were usually from only a single platform, such as Android, Windows Mobile, Symbian, or BlackBerry. In some cases, iOS devices at the time had limitations regarding background measurements and the release of applications in the app store \cite{bohmer2011}.

\subsubsection{Multidevice Usage Studies}
Multidevice studies can be further subclassified by the utilized data collection method or methods.

\paragraph{Interviews and Direct Observation}
Many early multidevice studies employed interviews or direct user observation. These methods are less scalable and accurate than device based monitoring but can provide contextual information such as why a user switches between devices during specific activities.

Dearman and Pierce \cite{dearman2008} examined multidevice usage of a panel of industry and academic users through semi-structured interviews. They highlighted, for example, that users use multiple devices for a variety of reasons including differences in device form factor and portability. Though tablet devices were not yet widely used at the time of the study. Similarly, Oulasvirta and Sumari \cite{oulasvirta2007} and Santosa and Wigdor \cite{santosa2013} also performed semi-structured interviews and user observation studies of multidevice users and workflows in the workplace context. Santosa and Wigdor \cite{santosa2013} especially highlighted the inadequacies of current methods for handing parallelism and data fragmentation in multidevice use cases.

Finally, Matthews et al. \cite{matthews2009} performed semi-structured interviews of industry smartphone users and examined the interplay of smartphone and PC usage. They found that, in contrast to basic cellphones, smartphones prompt both novel usage and substituted usage from PCs as functionality between smartphones and PCs partially overlaps. Furthermore, they report that smartphone attributes such as low initial access time gave smartphones an advantage over PCs for certain tasks in certain contexts (such as checking email). Thus their results are similar to the device substitution we explore between smartphone and tablet devices (wherein, for example, significant mobile game usage is substituted to larger screen tablets).

\paragraph{User Completed Electronic Diaries}
Other multidevice studies employed user completed electronic diaries in which a user tracks their own multidevice usage. This method is more scalable than interview or observational methods but still less accurate than device based monitoring as the approach relies on users, that are typically susceptible to recall bias \cite{dereuver2012}, to complete the diaries. However, similar to the interview method, user completed diaries can collect contextual information such as why a user performs a specific task.

Muller et al. \cite{muller2015} analyzed smartphone and tablet usage of a user panel through user completed electronic usage diaries. Similar to our study, their panel includes users that tracked both their smartphone and tablet device usage. They also report on the types and frequencies of secondary activities performed during tablet and smartphone use. Similarly, Google \cite{google2012} used usage diaries and surveys to study the multidevice usage of a large US based panel. The research found that over 90\% of panelists reported starting a task on one device and continuing the task on another device. However relatively few panelists (between 3-8\% depending on activity) reported starting an activity on a smartphone and later continuing on a tablet or the reverse sequence. 

\paragraph{Combinations of Methods and Other Methods}
Finally, a few studies have used a combination of these methods or examined multidevice usage from a service provider rather than user centric viewpoint.

Jokela et al. \cite{jokela2015} investigated multidevice usage of a small Finland based panel through both electronic diary and interview methods. They included a broad range of devices including tablets and, for example, home media centers in their multidevice usage definition. The investigation found the users reported significant sequential and parallel related but little parallel non-related multidevice usage, potentially because users do not easily realize non-related parallel usage.  Contrastingly, device based monitoring will always detect parallel usage but classification into related or non-related categories requires additional information and analysis. Similarly, Microsoft \cite{microsoft2013} also used electronic diary and interview methods to study multidevice usage. The research studied a mix of smartphones, tablets, e-readers, gaming consoles and laptops and classified so called multi-screening behavior across these devices as sequential, simultaneous or separate.

Montanez et. al \cite{montanez2014} and Wang et al. \cite{wang2013} studied cross device search in the context of a major US web search provider. Wang et al. found, for example, that about 15\% of device switching between searching involves contiguous tasks, in other words sequential multidevice sessions. 

Hintze et al. \cite{hintze2014} analyzed both smartphone and tablet usage through device based monitoring via the large and well known Device Analyzer dataset. Device Analyzer \cite{wagner2014} is a large scale dataset based on a popular Android monitoring app. The dataset contains both smartphone and tablet devices however correlating ownership of smartphone and tablet devices to a single individual is not possible thus limiting multidevice session analysis. Furthermore individual application names are not available, thus further limiting analysis.

\subsubsection{Multidevice Enablers and Frameworks}
A logically related and currently very active area of study is the building of enablers and frameworks for more convenient multidevice usage.

Taivalsaari et al. \cite{taivalsaari2014} discuss liquid software as a general manifesto of guidelines that multidevice interaction should follow. Hamilton and Wigdor \cite{hamilton2014} present Conductor; a framework for the construction of multidevice applications, whereas Desrulle and Gielen \cite{desruelle2014} discuss the high level architectures of such systems in the context of web applications. Finally, Schreiner et al. \cite{schreiner2015} present a versatile web based multidevice app framework that is agnostic to, for example, the underlying device to device communication method. Relationally, our research looks to help find potentially useful (app) targets for such enablers and to help describe the current state of multidevice usage so the value of these enablers can be estimated.

\subsection{Allen's Time Interval Relations} \label{sect_allen_relations}
\begin{figure}[!t]
\centering
\includegraphics[width=6.4in]{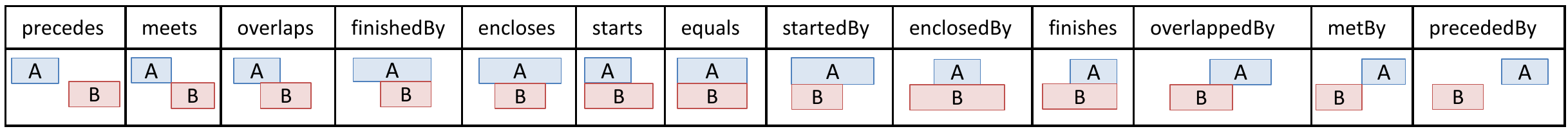}
\caption{Allen's Time Interval Relations}
\label{fig_allen}
\end{figure}

Allen's thirteen basic time interval relations are the building blocks of Allen's Interval Algebra \cite{allen1983}. The relations are {\it distinct} (a pair of intervals are described by one and only one relation), {\it exhaustive} (any pair of intervals can be described by one of the relations) and {\it qualitative} (no numeric time spans are required). Figure \ref{fig_allen} shows all of the thirteen basic relations that two finite time intervals {\it a} and {\it b} can have. In the figure, time is expressed from left to right. So, for example, in the first cell interval {\it a} starts and ends before interval {\it b} starts. In the second cell interval {\it b} starts exactly as interval {\it a} ends. Intuitively, six of the relations can be described as converses of six other relations. For example, the converse of {\it a} {\bf meets} {\it b} is {\it b} {\bf metBy} {\it a}.

In this paper we often use the notion of a session. In the context of, for example, smartphone usage an app session can be defined as a period of application usage with a clear start and end time. This type of session is equivalent to Allen's notion of a finite time interval and thus we can use Allen's relations in defining and examining multidevice usage. The relations formalize the temporal relationships between (usage) sessions of one or more devices. 

This work is the first multidevice paper, as far as we know, to exploit such a formalization and thus to bring this foundation into the multidevice usage area.

\section{Datasets} \label{datasets}
The two primary datasets are subsets of a large United States based user panel organized by Verto Analytics\footnote{http://vertoanalytics.com/} in February 2015. Panelists were recruited online and were given an initial recruitment survey that asked about the devices they own. Panelists were instructed to install custom monitoring applications to all of their applicable devices (in other words, the devices they reported owning including smartphone, tablet, and PC). The monitoring applications logged events such as an app moving to the foreground or background of the device. Only panelists that installed the monitoring applications to all their applicable devices were considered for the panel. All panelists were paid for participation. All analyzed user data was anonymized with no personally identifying information.

The first dataset (hereafter dataset 1) consists of one month (February 2015) of smartphone and tablet application usage data from the 65 panelists with both a smartphone and tablet device. Only active panelists were considered valid and included in the subset. We define a panelist as active if for both of their devices the length between their first usage of the month and last usage of the month is at least 23 days. The subset includes panelists with devices from both major mobile platforms: Apple iOS and Google Android. Specifically, 46 panelists had both an Android smartphone and tablet, 10 panelists had both an iOS smartphone and tablet, 5 panelists had an iOS smartphone and Android tablet, and 4 panelists had Android smartphone and iOS tablet.

The second dataset (hereafter dataset 2) consists of one month (February 2015) of smartphone application usage data from 496 panelists with a smartphone but not a tablet device. In other words, these panelists reported on their panel recruitment survey that they do not own a tablet device. Again only active panelists were considered valid and included in the subset and the subset includes both major mobile platforms. 402 panelists had Android smartphones and 94 panelists had iOS smartphones.

In terms of representativeness, the large panel was recruited with the purpose of obtaining a nationally representative panel and thus is relatively diverse. However all opt-in panels inherently use non-probability sampling and thus representativeness is a concern\footnote{We note that in contrast to panels were panelists take surveys, the device based monitoring collection method is not generally susceptible to, for example, false/fake answers, careless answers, or repeatedly giving the same answer.}. We refer to Hays et al. \cite{hays2015} for further discussion. Furthermore, as mentioned, the recruitment process utilized a recruitment survey that screened potential panelists to improve the demographic and technographic match between the accepted panelists and the population (known as a quota-sampling approach).

For reference we provide a summary of panelist demographic and technographic data for each dataset along with demographic data for US smartphone users in general in Table \ref{panel_demos}. The clearest demographic discrepancies are that the datasets over represent females and users with lower household incomes. Thus these factors should be considered when generalizing the results. We discuss our overall view of generalizability in panel based studies in Section \ref{discussion}.

In terms of the types of monitored device types, personal computer usage data was also collected from panel users but considered out of scope for this paper. Though analysis of PC usage in a multidevice context would also be valuable, a significant issue is that the simple foreground app session definition is not directly transferable to the personal computer context. For example, PC users often have multiple viewable app windows open simultaneously. Furthermore, the limited mobility of laptops and desktops means that these devices often remain continuously connected to an AC socket, thus users can often leave their screens on even when not interacting with the device (for example by disabling sleep mode). Therefore, more advanced user activity detection would be needed.

\begin{table}[!t]
\scriptsize
\centering
\begin{threeparttable}
\caption{Demographic and Technographic Summaries of Datasets}
\label{panel_demos}
\begin{tabular}{lrrr}
\toprule
 & {\bf Dataset 1 } & {\bf Dataset 2 } & {\bf US Smartphone Users\tnote{a}}\\
{\bf Mean Age (Years)\tnote{b}} & 38.53 (11.91) & 37.06 (12.63) & 41.30 (15.08)\\
{\bf Gender (\% Male)} & 23.08 & 26.61 & 50.08\\ 
{\bf Education (\% w/ Some College or Less)} & 56.92 & 63.10 & 53.04\\ 
{\bf Marital Status (\% Married)} & 44.62 & 41.33 & 48.96\\ 
{\bf Household Income (\% \textless 50K USD)} & 47.69 & 65.93 & 40.72\\ 
{\bf Mean Household Size} & 2.97 (1.54) & 2.95 (1.50) & 3.05 (1.61)\\ 
{\bf Mean Children in Household} & 0.94 (1.06) & 0.93 (1.21) & 0.74 (1.27)\\ 
{\bf Race (\% White)} & 72.31 & 70.97 & 71.53\\
\midrule
{\bf Mean Smartphone Display Size (inches)} & 4.61 (0.71) & 4.58 (0.60) & -\\
{\bf Mean Smartphone Display Density (ppi)\tnote{c}} & 358.00 (72.79) & 343.86 (84.67) & -\\
{\bf Mean Smartphone Total CPU Frequency (GHz)\tnote{d}} & 5.57 (3.33) & 5.50 (3.20) & -\\
{\bf Mean Time Since Smartphone Release (years)\tnote{e}} & 2.06 (1.11) & 1.92 (1.04) & -\\
\midrule
{\bf Mean Tablet Display Size (inches)} & 8.59 (1.27) & - & -\\
{\bf Mean Tablet Display Density (ppi)\tnote{c}} & 203.89 (57.86) & - & -\\
{\bf Mean Tablet Total CPU Frequency (GHz)\tnote{d}} & 3.58 (2.04) & - & -\\
{\bf Mean Time Since Tablet Release (years)\tnote{e}} & 2.05 (0.86) & - & -\\
\bottomrule
\end{tabular}
\begin{tablenotes}
    \scriptsize
    \item[a] US smartphone user demographic data is from Pew Research survey (June-July 2015, subpop with smartphone n=1327) \cite{pew2015}. The survey utilizes weighting to population parameters of census data to create nationally representative results (refer to \cite{pew2016}). We note that Verto Analytics also performs its own national surveys, we utilize the Pew Research survey only for brevity.
    \item[b] All mean values also include standard deviations
    \item[c] In pixels per inch
    \item[d] Total frequency of CPU as number of cores multiplied by frequency per core
    \item[e] Time from approximate release month of device model until Feb 2015
\end{tablenotes}
\end{threeparttable}
\end{table}

\section{Usage session} \label{usage_session}
\subsection{Session Definitions} \label{sect_session_defs}
In order to study mobile usage sessions we first define several important concepts we will use throughout the analysis. The definitions are partly based on Allen's time interval relations described in section \ref{sect_allen_relations}. First, we define an {\it app session} as a time interval starting with an application moving to the foreground of the device and ending with the application moving out of the foreground (either replaced by a different app or screen off). This app session definition has been used in previous literature \cite{bohmer2011,falaki2010,soikkeli2013}. Next, we define a {\it single device usage session} as a collection of app sessions on a single device with a maximum timeout value between two sequential app sessions. This definition has also been used previously \cite{bohmer2011,soikkeli2013}, but in comparison to previous work, the Allen's relations approach we adapt to usage session construction (see section \ref{sect_session_construction}) offers a more solid foundation for characterizing usage sessions and their relations, and more importantly can be easily extended to multidevice usage.

Thus we describe our definitions in terms of Allen's relations. Since by definition single device app sessions are sequential, only {\bf precedes} and {\bf meet} of Allen's relations (and their respective converse relations) are applicable. If two app sessions {\bf meet}, the sessions will always belong to the same usage session. If an app session {\bf precedes} or is {\bf precededBy} another app session within the maximum timeout, then the app session will also belong to the same usage session. Henceforth, we denote this relation as {\bf precedes (precededBy) within time-window} (abbreviated as {\bf precedes (precededBy) within TW}). Thus through a combination of these four relations multiple {\it app sessions} can be grouped into {\it single device usage sessions}. 

\begin{figure*}[!t]
\centering
\includegraphics[width=6.4in]{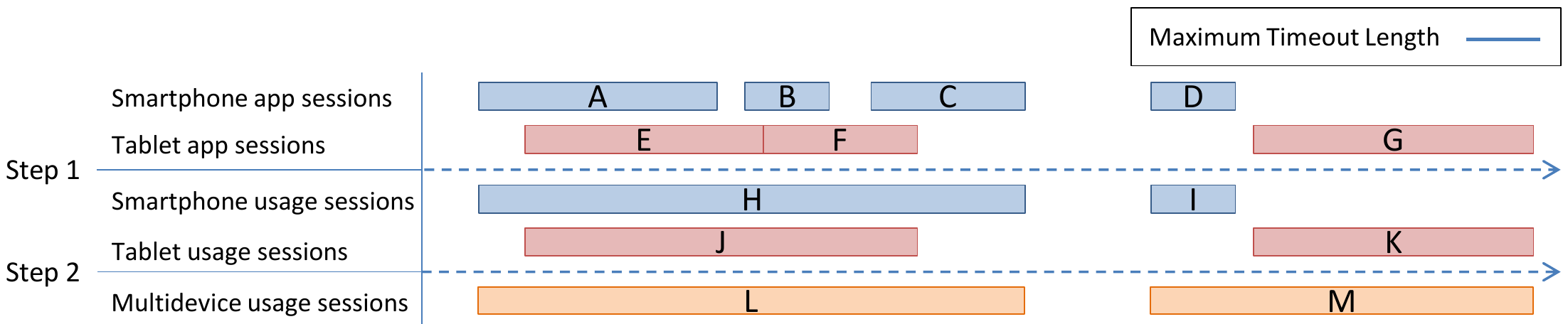}
\caption{Multidevice usage session construction - Step 1: Apply usage session definition to each device's app sessions (merge A,B,C to H and E,F to J; D to I and G to K are degenerate single app session cases), 
Step 2: Collapse simultaneous, meeting and preceding within time-window usage sessions into multidevice usage sessions (merge H,J to L and I,K to M).
}
\label{fig_sessions}
\end{figure*}

For multiple devices (of a single user in our case) all of Allen's relations are applicable. As mentioned {\bf precedes} and {\bf meet} (with their respective converse relations) describe sequential usage, while the remaining relations describe (at least partly) simultaneous usage of the devices. For constructing a {\it multidevice usage session} we follow a two-step process. First, we apply the single device usage session definition separately for each device. Second, we examine the relations of the usage sessions of the multiple devices. If the single device usage sessions are (at least partly) simultaneous, {\bf meet} (are {\bf metBy}) or {\bf precede} (are {\bf precededBy}) {\bf within TW} then they belong to the same {\it multidevice usage session}. Figure \ref{fig_sessions} illustrates this process with smartphone and tablet sessions (two devices).

Finally, we define several descriptive session types that we will use throughout the paper. Specifically, each single device usage session (and by extension each app session contained within a session) can be classified along two distinct binary dimensions: device type ({\it smartphone} or {\it tablet}) and containment in a multidevice usage session (if contained in a multidevice session then {\it mixed} or if not then {\it pure}).

\subsection{Maximum Timeout Value}
In both the single and multidevice cases, the maximum timeout value (used in session construction) is an important parameter that affects the eventual interpretation of the results. Ideally the timeout value should be a moderate value that accurately reflects how users psychologically divide their daily device usage. However, as mentioned in section \ref{sect_rel_work}, various authors have used different values with no best practice prevailing, therefore we briefly examine the effects of a range of different timeout values.

\begin{figure*}[!t]
\centering
\includegraphics[width=5.5in]{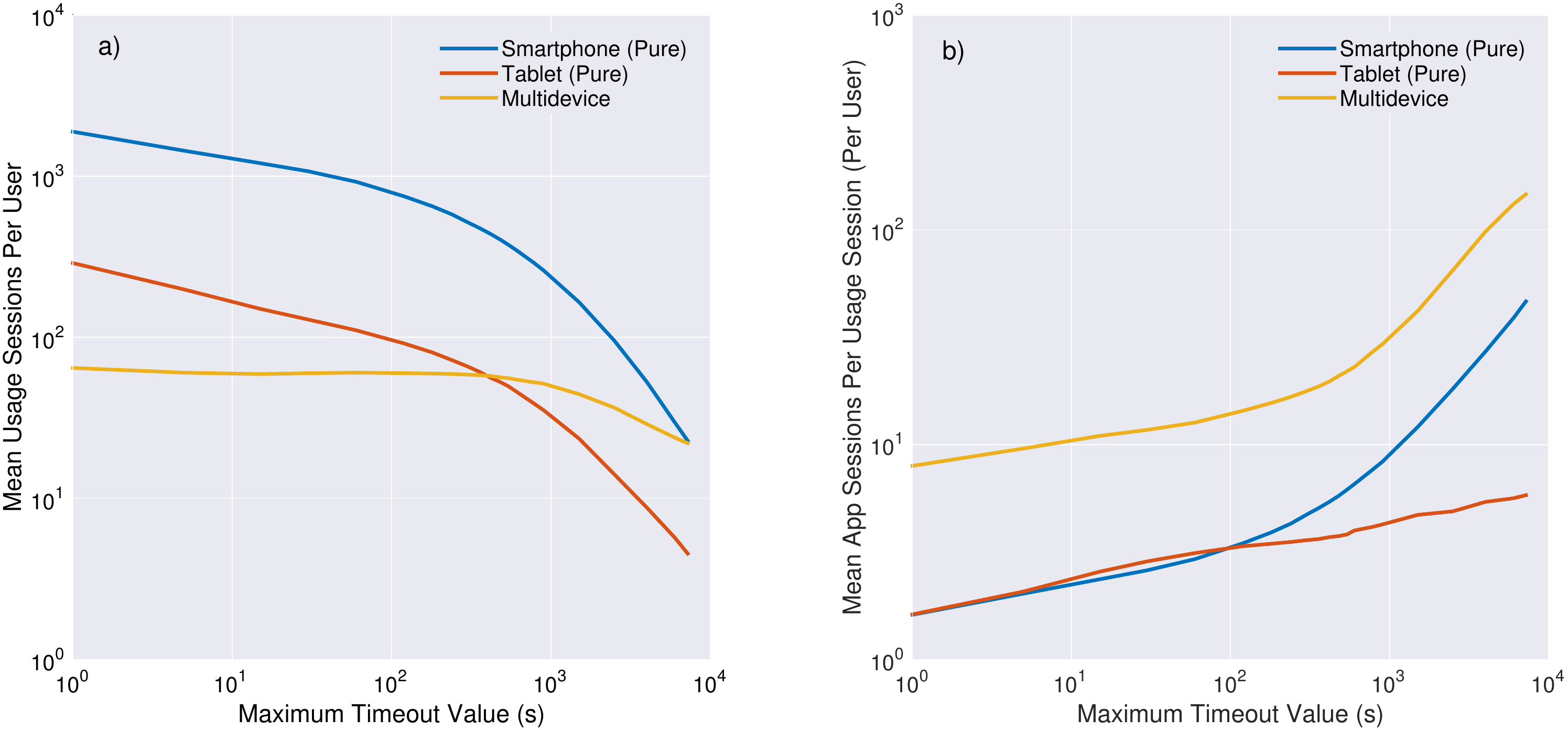}
\caption{a) Mean Usage Sessions Per User by Timeout Window, b) Mean App Sessions Per Usage Session (Per User) by Timeout Window}
\label{timeout_window}
\end{figure*}

Figure \ref{timeout_window}a details the mean number of different types of usage sessions per user as a function of timeout value. Interestingly, the mean number of multidevice sessions is relatively stable for timeout values of 1-1000s, indicating that in this range multidevice sessions are primarily subsuming time-adjacent pure smartphone and tablet usage sessions. Whereas after 1000s, the significant decrease in multidevice sessions suggests that time-adjacent multidevice sessions themselves are merged. Figure \ref{timeout_window}b illustrates the mean app sessions per usage session per user as a function of timeout value. The increasing number of app sessions per usage session further supports the previous indications.

In terms of selecting a practical timeout value for analysis purposes, we select a mid-range compromise value of 60s that preserves a large number of pure smartphone and tablet usage sessions and does not indicate significant merging of multidevice sessions. Furthermore, we use the same value for creating both single device usage sessions and multidevice usage sessions. Though an argument could be made that switching devices in sequential multidevice sessions might often take longer than 60 seconds and thus a longer timeout value could be used. However, we leave this analysis for future work.

\subsection{Usage Session Construction} \label{sect_session_construction}
In order to analyze smartphone, tablet and multidevice usage on a session level we first construct the usage sessions. We follow the session definitions of section \ref{sect_session_defs}. Table \ref{tab_usage_session_stats} details statistics related to the usage session construction process of dataset 1. 

In the single device case one app session can have from zero to two relations: no relation to a preceding or following app session, a single relation to a preceding or a following app session, or relations to both preceding and following app sessions. For Table \ref{tab_usage_session_stats} we count all these relations. Table \ref{tab_a1} in Appendix A details similar statistics for dataset 2.

Table \ref{tab_usage_session_stats} also breaks down the types of Allen's relations observed between the smartphone and tablet usage sessions contained in multidevice usage sessions. The relations are written with the smartphone session preceding the relation and the tablet session following. In other words, we can read the statistics as follows: smartphone session [relation] tablet session. In the majority of cases (over 61 \%) a smartphone session is {\bf enclosedBy} a tablet session. The second most frequent case (with 11 \%) is that a smartphone session {\bf encloses} a tablet session. A majority of the rest of the cases fall relatively evenly between {\bf overlaps} and {\bf precedes within TW} and their converses.

\begin{table}[!t]
\scriptsize
\centering
\begin{threeparttable}
\caption[Usage session construction statistics]{Usage session construction statistics}
\label{tab_usage_session_stats}
\begin{tabular}{@{}lrrr@{}}
\toprule
 & {\bf Smartphone usage} & {\bf Tablet usage} & {\bf Multidevice usage} \\ \midrule
{\bf Number of app sessions}                 & 330523            & 63239         & 97620           \\
{\bf Number of single device usage sessions} & 78798            & 12708          & 18136           \\
{\bf Number of multidevice usage sessions}   & -                          & -                      & 4517             \\
\bottomrule
{\bf Share of Allen's relations}             &                            &                        &    \\
precedesWithinTW                             & 8.16 \%           & 4.65 \%       & 6.16 \%        \\
meets                                        & 42.07 \%        & 45.44 \%     & 0.10 \%        \\
overlaps                                     & -                          & -                      & 8.50 \%       \\
finishedBy                                   & -                          & -                      & 0.04 \%        \\
encloses                                     & -                          & -                      & 10.70 \%      \\
starts                                       & -                          & -                      & 0.12 \%       \\
equivalent                                   & -                          & -                      & 0.00 \%        \\
startedBy                                    & -                          & -                      & 0.11 \%        \\
enclosedBy                                   & -                          & -                      & 61.40 \%     \\
finishes                                     & -                          & -                      & 0.11 \%        \\
overlappedBy                                 & -                          & -                      & 7.40 \%      \\
metBy                                        & 41.60 \%         & 45.26 \%     & 0.11 \%        \\
precededByWithinTW                           & 8.16 \%           & 4.65 \%       & 5.27 \%      \\ \bottomrule
\end{tabular}
\end{threeparttable}
\end{table}

\section{Results} \label{results}
In this section, we present and discuss the main results of our analysis and where appropriate compare these results to prior work.
\subsection{Basic Descriptive Statistics}
\subsubsection{Dataset Level Statistics}
In this section we study the dataset level basic characteristics of the single device and multidevice usage sessions. In other words, the statistics describe the aggregate of all sessions from all users of dataset 1. Table \ref{tab_basic_stats_whole} details descriptive statistics related to usage session lengths and the number of app sessions per usage session (for dataset 2 see Table \ref{tab_a2} in Appendix A). We calculate the statistics separately for all smartphone usage sessions (combined pure and mixed), all tablet usage sessions (combined pure and mixed), pure smartphone usage sessions, pure tablet usage sessions and multidevice usage sessions. 

On general level, we find that tablet usage sessions are significantly longer than smartphone usage sessions but that the number of app sessions in a single device usage session is similar for both smartphones and tablets. Whereas, multidevice usage sessions are about 10 times longer and have about 5 times as many app sessions as compared to the single device usage sessions.

\begin{figure*}[!t]
\centering
\includegraphics[width=6.4in]{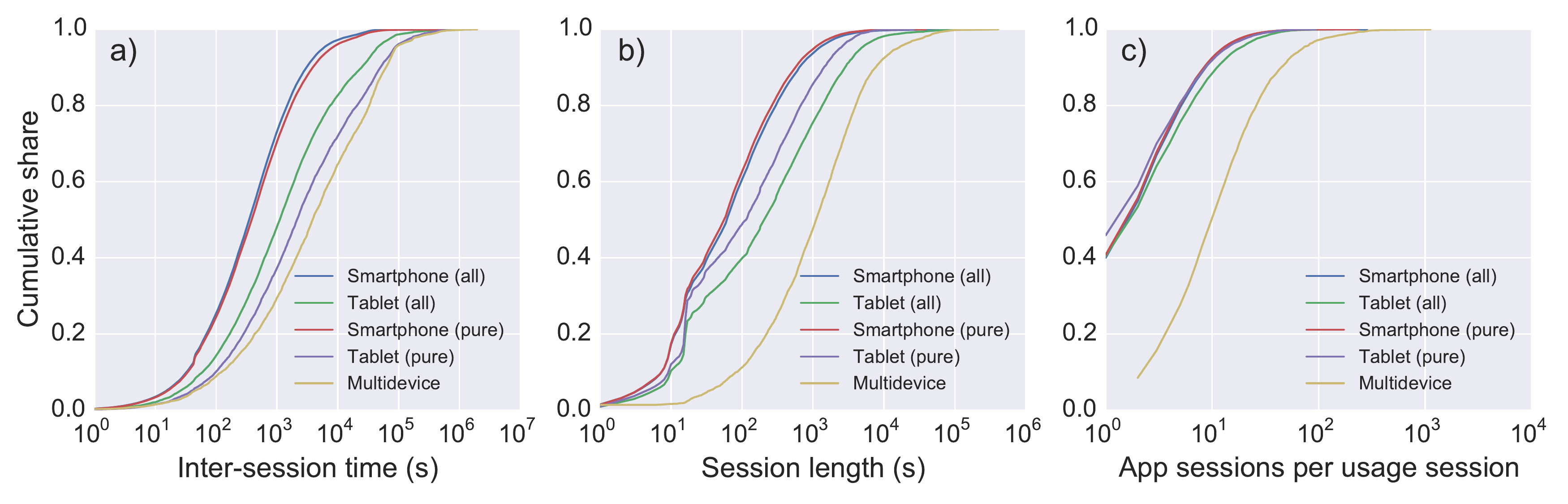}
\caption{Cumulative distributions (the inter-session time axis is shifted by 60s for readability)}
\label{fig_cdfs}
\end{figure*}

These reported mean and median values are indicative of the differences between sessions types but do not illustrate the full distributions very well. Figures \ref{fig_cdfs}b and  \ref{fig_cdfs}c show empirical distribution functions of usage session lengths and number of app sessions per usage session, respectively, for the different types of usage sessions. In terms of session lengths, the distributions show clear peaks for smartphone and especially for tablet sessions. A small peak appears at 10s and is associated with certain Samsung devices' touch user interface (known as TouchWiz). A larger more noticeable peak appears at around 15-17s and is associated with iOS applications not properly identified by the monitoring application. The difference in the share of these particular iOS sessions between pure and mixed tablet sessions is the main reason for the relatively large difference in session length values between these types. For brevity, we leave the further examination of these app specific anomalies for future research.

In terms of number of app sessions, only 8\% of all multidevice usage sessions have exactly two app sessions (the minimum for multidevice usage sessions by definition) while half of multidevice usage sessions have more than 10 app sessions. Thus app and interaction designers likely need to consider complex use cases of multiple devices well beyond the degenerate case of two simple sequential or simultaneous app sessions (as these degenerate cases are rare in practice). We further illustrate the wide range of multidevice app usage patterns in Section \ref{sect_session_patterns}. 

Table \ref{tab_basic_stats_whole} also shows how the usage is divided among the session types based on different definitions of usage including number of app sessions, number of usage sessions, and interaction time. Most interestingly when the division is based on interaction time, multidevice and single device usage divide almost equally with 50.9\% multidevice usage and 49.1\% single device usage. In other words, over half of all interaction time is considered multidevice usage. This reinforces the significant prevalence of multidevice usage for multidevice users and the potentially large impact of optimizations for multidevice usage.

If we examine the device types carefully, we find that a tablet usage session is much more likely to be part of a multidevice usage session, when compared to smartphone usage sessions. In fact, over 90\% of tablet interaction time and 93\% of tablet usage sessions are associated with multidevice usage. Comparatively, Google \cite{google2012} survey results, for example, indicate only 35\% of tablet sessions also include smartphone usage. Similarly, Muller et al. \cite{muller2015} reports device multitasking during tablet usage in only a small fraction (3.4\%) of user reported usage incidents. Though these studies are not fully comparable, the large differences in results still reinforce the importance of utilizing device monitoring methods in addition to self reporting methods.

\begin{figure*}[!t]
\centering
\includegraphics[width=6.4in]{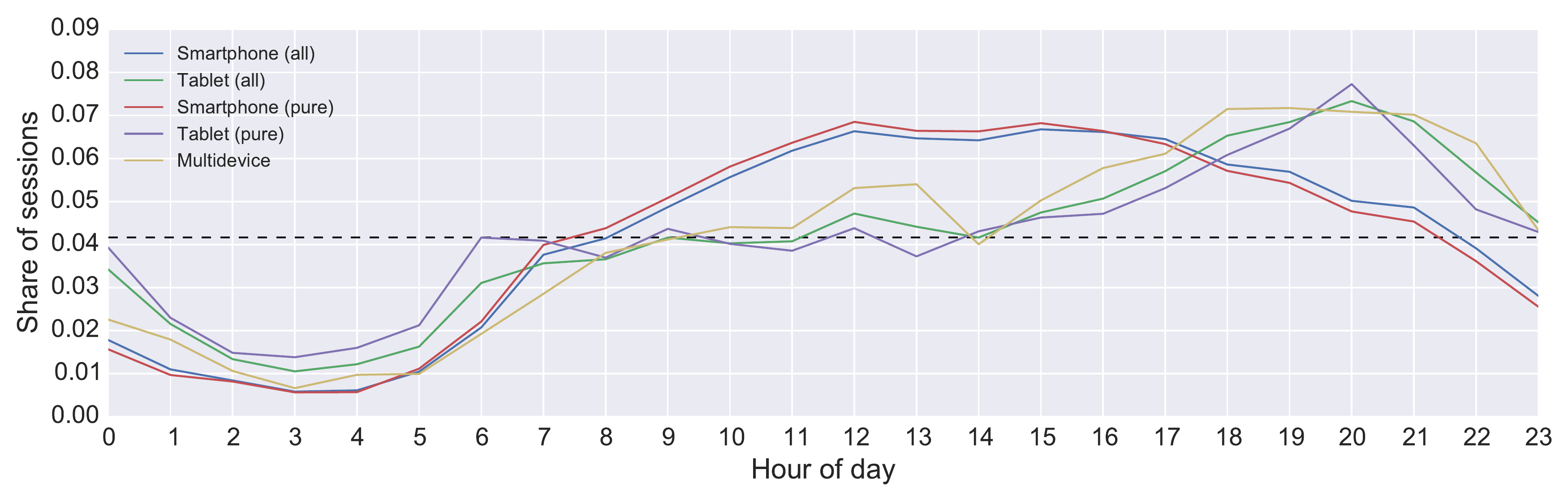}
\caption{Hourly Distributions}
\label{fig_hourly_dist}
\end{figure*}

Finally, in terms of temporal distributions, Figure \ref{fig_hourly_dist} shows hourly distributions of the different types of usage sessions. Smartphone usage is distributed relatively uniformly across the office hours with a small shift towards the evening.  Whereas tablet and multidevice usage is focused in the late evening. These distributions support the notion of smartphones being the ubiquitous all-day work horses of mobile device usage, while tablets are related to leisure time and the home context. Furthermore, the distributions also support and validate previous work based on self reported diaries such as \cite{muller2015} that illustrate similar distributions.

\begin{table*}[!t]
\scriptsize
\centering
\begin{threeparttable}
\caption[Basic statistics on the whole dataset level]{Basic statistics on the whole dataset level\tnote{a}}
\label{tab_basic_stats_whole}
\begin{tabular}{@{}lrrrrr@{}}
\toprule
 & {\bf \begin{tabular}[c]{@{}r@{}}Smartphone - all\\ N = 78798\\ \end{tabular}} & {\bf \begin{tabular}[c]{@{}r@{}}Tablet - all\\ N = 12708\\ \end{tabular}} & {\bf \begin{tabular}[c]{@{}r@{}}Smartphone - pure\\ N = 63930\\\end{tabular}} & {\bf \begin{tabular}[c]{@{}r@{}}Tablet - pure\\ N = 6871\\ \end{tabular}} & {\bf \begin{tabular}[c]{@{}r@{}}Multidevice\\ N = 4517\\ \end{tabular}} \\ \midrule
{\bf Usage session length (s)} & \multicolumn{1}{l}{} & \multicolumn{1}{l}{} & \multicolumn{1}{l}{} & \multicolumn{1}{l}{} & \multicolumn{1}{l}{} \\
mean & 312 & 1461 & 245 & 506 & 4241 \\
median & 61 & 197 & 56 & 114 & 1147 \\
std. dev. & 1509 & 8154 & 829 & 1074 & 14284 \\
{\bf \begin{tabular}[c]{@{}l@{}}App sessions per usage\\ session\end{tabular}} & \multicolumn{1}{l}{} & \multicolumn{1}{l}{} & \multicolumn{1}{l}{} & \multicolumn{1}{l}{} & \multicolumn{1}{l}{} \\
mean & 4.13 & 4.95 & 3.96 & 3.97 & 21.61 \\
median & 2.00 & 2.00 & 2.00 & 2.00 & 10.00 \\
std. dev. & 6.46 & 7.92 & 5.90 & 5.98 & 45.47 \\
{\bf Percentage of all usage} & \multicolumn{1}{l}{} & \multicolumn{1}{l}{} & \multicolumn{1}{l}{} & \multicolumn{1}{l}{} & \multicolumn{1}{l}{} \\
based on app sessions & 83.5 \% & 16.5 \% & 67.0 \% & 7.2 \% & 25.8 \% \\
based on usage sessions & 85.9 \% & 14.1 \% & 84.9 \% & 9.1 \% & 6.0 \% \\
based on interaction time & 55.8 \% & 44.2 \%  & 40.0 \% & 9.2 \% & 50.9 \% \\ \bottomrule
\end{tabular}
    \begin{tablenotes}
      \small
      \item[a] N is the number of usage sessions. Smartphone (all) and Tablet (all) indicates all smartphone and tablet usage or app sessions, whereas Smartphone (pure) and Tablet (pure) indicates only pure smartphone and tablet usage or app sessions.
    \end{tablenotes}
\end{threeparttable}
\end{table*}

\subsubsection{Per User Level Statistics}
In the previous section we examined the usage sessions on the entire dataset level, in other words all sessions of all users were aggregated. Table \ref{tab_basic_stats_user} shows similar statistics, again for dataset 1, but based on a per user level (for statistics for dataset 2 see Table \ref{tab_a3} in Appendix A). More specifically, the session lengths and number of app sessions per usage session are based on per user medians and the usage sessions per day and interaction time per day are themselves per user statistics.

The per user results reveal considerable diversity across users also in the multidevice usage case. Session length medians show an order of magnitude difference between users and in the case of app sessions per usage session from 5 to 10 fold differences. Thus we extend to the multidevice context previous studies, for example \cite{hintze2014}, that report significant diversity in usage across users.

In terms of per user interaction time, we find that typically no single device dominates but users spend considerable time with both their smartphone and tablet. This relatively equitable time distribution suggests that mobile ad inventories per user will also be roughly equitable depending on ad refresh rates. Thus informing multidevice mobile advertising strategies.

Furthermore, the overall interaction time with both smartphones and tablets is considerably higher than previous studies\footnote{For smartphones \cite{hintze2014} reports a median daily usage of 82 mins, compared to our median daily usage of 261 mins for dataset 2. For tablets \cite{hintze2014} reports a median daily usage of 67 mins, compared to our median daily usage of 115 mins for dataset 1.} (see Table 1 in \cite{hintze2014}). These differences are potentially due to increased pervasiveness of these devices over time, differences in the types of users between studies, or a combination of these factors. For reference, the dataset from \cite{hintze2014} has users from over 175 countries whereas our dataset has only US users.

\begin{table*}[!t]
\scriptsize
\centering
\begin{threeparttable}
\caption[Basic statistics on per user level]{Basic statistics on per user level\tnote{a}}
\label{tab_basic_stats_user}
\begin{tabular}{@{}lrrrrr@{}}
\toprule
 & {\bf Smartphone - all} & {\bf Tablet - all} & {\bf Smartphone - pure} & {\bf Tablet - pure} & {\bf Multidevice} \\ \midrule
{\bf Usage session length (s)} & \multicolumn{1}{l}{} & \multicolumn{1}{l}{} & \multicolumn{1}{l}{} & \multicolumn{1}{l}{} & \multicolumn{1}{l}{} \\
mean & 85 & 463 & 78 & 211 & 2041 \\
median & 57 & 215 & 54 & 121  & 1063 \\
std. dev. & 109  & 623  & 99 & 259 & 2983 \\
{\bf \begin{tabular}[c]{@{}l@{}}App sessions per usage\\ session\end{tabular}} &  &  &  &  &  \\
mean & 2.28 & 2.73 & 2.25 & 2.15 & 13.03 \\
median & 2.00 & 2.00 & 2.00 & 2.00 & 11.00 \\
std. dev. & 1.04  & 2.00 & 1.00 & 1.42 & 8.95 \\
{\bf Usage sessions per day} &  &  &  &  &  \\
mean & 44.5 & 7.15 & 36.05  & 3.85 & 2.59 \\
median & 43.13 & 5.01 & 34.3 & 2.82 & 1.79 \\
std. dev. & 22.17 & 6.78 & 19.96 & 4.6 & 2.38 \\
{\bf \begin{tabular}[c]{@{}l@{}}Interaction time per day\\ (min)\end{tabular}} &  &  &  &  &  \\
mean & 234.21 & 174.95 & 146.69 & 32.57 & 184.22 \\
median & 213.86 & 115.01 & 114.78  & 19.23 & 121.80 \\
std. dev. & 167.89 & 188.12 & 97.95 & 44.37 & 206.05 \\
\bottomrule
\end{tabular}
    \begin{tablenotes}
      \small
      \item[a] Usage session length and App sessions per usage session statistics are calculated from the per user medians of these metrics. Smartphone (all) and Tablet (all) indicates all smartphone and tablet usage sessions, whereas Smartphone (pure) and Tablet (pure) indicates only pure smartphone and tablet usage sessions.
    \end{tablenotes}
\end{threeparttable}
\end{table*}

\subsection{Patterns within Multidevice Usage Sessions} \label{sect_session_patterns}
In this section, we study the temporal patterns within multidevice usage sessions. In other words, the configurations of the device usage sessions that form each multidevice usage session. 

First, to quantify the temporal similarity between two multidevice usage sessions we need a similarity metric. We use an interval sequence similarity measure known as Interval-Based Sequence Matching (IBSM) \cite{kotsifakos2013}. The measure is based on representing sequences of finite intervals as matrices. The similarity between these matrices is then the Frobenius norm of the difference between the matrices. In our case, each matrix represents a multidevice usage session, each matrix row represents a specific device (in our case always two rows for smartphone and tablet, with smartphone always the first row and tablet the second) and each matrix element represents the binary usage state (active or inactive) at a discrete time (a second). In other words, a multidevice usage session with a smartphone session from $t=1$ to $t=4$ and a tablet session from $t=3$ to $t=5$ would be represented as:
\vspace{3mm}

{\centering
$\begin{bmatrix}
1 & 1 & 1 & 1 & 0 \\ 
0 & 0 & 1 & 1 & 1
\end{bmatrix}$\par
}
\vspace{3mm}
In cases where the matrices to be compared are of different lengths then each row of the shorter matrix is re-sized through linear interpolation.

The IBSM measure is used instead of Allen's relations because while Allen's relational algebra is a fundamental tool for understanding interval relations, the categorical relations are quite strict. In other words, only a small difference in position of a session relative to another session can result in a totally different relation, even though the two configurations might be considered in a sense similar.

With the similarity measure, we divide the multidevice usage sessions into groups based on the similarity of each session's matrix to well defined artificial prototype matrices. Specifically, we use all 256 configurations of the size $2\times4$ (row $\times$ column) binary matrix as prototypes and each prototype represents a group (thus we have 256 distinct groups). Importantly since we are representing relatively large matrices, on average 2041 columns, by small matrices, always 4 columns, these prototype groups are rough highly aggregated approximations. We refer to each group by the integer number represented by the binary of the prototype matrix when the second row is concatenated after the first.

The choice of 4-column prototype matrices (rather than for example larger or small prototype matrices) relates to a trade-off over the level of detail that proves useful in understanding common patterns. Specifically, large prototype matrices (for example the same size as the actual matrices) would just represent every actual matrix by it's own prototype, whereas, very small prototype matrices (1 column) would just obfuscate most of the interesting differences between the matrices. Thus the 4-column size represents a good compromise size that maintains a useful level of detail.

Table \ref{md_prototype_groups} details the ten most frequent groups of all multidevice usage sessions. Interestingly, we find that the top two groups represent primarily usage of a single device over the length of the session along with sparse usage of a second device. The relatively high frequency of group 15 is expected given that tablet usage sessions are generally longer and smartphone usage sessions are shorter and burstier. However surprisingly group 240 (the bit-wise complement of group 15) is also quite frequent. 

\begin{table}[!t]
\caption{Most Frequent Prototype Groups of Multidevice Usage Sessions (All Sessions and Per User)}
\label{md_prototype_groups}
\centering
\begin{tabular}{@{}ccccccc@{}}
\toprule
\multicolumn{3}{c}{All Multidevice Usage Sessions} & & \multicolumn{3}{c}{Multidevice Usage Sessions Per User} \\
Group Number & Group Matrix & Frequency && Group Number & Group Matrix & Frequency\\ 
\midrule
\addlinespace[0.5em]
15 & $\begin{bmatrix}
0 & 0 & 0 & 0 \\ 
1 & 1 & 1 & 1 
\end{bmatrix}$ & 24.07\% && 
15 & $\begin{bmatrix}
0 & 0 & 0 & 0 \\ 
1 & 1 & 1 & 1 
\end{bmatrix}$ & 25.70\% \\
\addlinespace[0.5em]
240 & $\begin{bmatrix}
1 & 1 & 1 & 1 \\ 
0 & 0 & 0 & 0 
\end{bmatrix}$ & 10.90\% && 
240 & $\begin{bmatrix}
1 & 1 & 1 & 1 \\ 
0 & 0 & 0 & 0 
\end{bmatrix}$ & 12.95\% \\
\addlinespace[0.5em]
135 & $\begin{bmatrix}
1 & 0 & 0 & 0 \\ 
0 & 1 & 1 & 1 
\end{bmatrix}$ & 4.45\% &&
135 & $\begin{bmatrix}
1 & 0 & 0 & 0 \\ 
0 & 1 & 1 & 1 
\end{bmatrix}$ & 3.62\% \\
\addlinespace[0.5em]
143 & $\begin{bmatrix}
1 & 0 & 0 & 0 \\ 
1 & 1 & 1 & 1 
\end{bmatrix}$ & 3.72\% &&
30 & $\begin{bmatrix}
0 & 0 & 0 & 1 \\ 
1 & 1 & 1 & 0 
\end{bmatrix}$ & 3.37\% \\
\addlinespace[0.5em]
30 & $\begin{bmatrix}
0 & 0 & 0 & 1 \\ 
1 & 1 & 1 & 0 
\end{bmatrix}$ & 2.90\% && 
143 & $\begin{bmatrix}
1 & 0 & 0 & 0 \\ 
1 & 1 & 1 & 1 
\end{bmatrix}$ & 3.32\% \\
\addlinespace[0.5em]
79 & $\begin{bmatrix}
0 & 1 & 0 & 0 \\ 
1 & 1 & 1 & 1 
\end{bmatrix}$ & 2.50\% && 
79 & $\begin{bmatrix}
0 & 1 & 0 & 0 \\ 
1 & 1 & 1 & 1 
\end{bmatrix}$ & 2.60\% \\
\addlinespace[0.5em]
195 & $\begin{bmatrix}
1 & 1 & 0 & 0 \\ 
0 & 0 & 1 & 1 
\end{bmatrix}$ & 2.24\% && 
31 & $\begin{bmatrix}
0 & 0 & 0 & 1 \\ 
1 & 1 & 1 & 1 
\end{bmatrix}$ & 2.36\% \\
\addlinespace[0.5em]
225 & $\begin{bmatrix}
1 & 1 & 1 & 0 \\ 
0 & 0 & 0 & 1 
\end{bmatrix}$ & 2.17\% && 
195 & $\begin{bmatrix}
1 & 1 & 0 & 0 \\ 
0 & 0 & 1 & 1 
\end{bmatrix}$ & 2.11\% \\
\addlinespace[0.5em]
31 & $\begin{bmatrix}
0 & 0 & 0 & 1 \\ 
1 & 1 & 1 & 1 
\end{bmatrix}$ & 2.17\% && 
225 & $\begin{bmatrix}
1 & 1 & 1 & 0 \\ 
0 & 0 & 0 & 1 
\end{bmatrix}$ & 1.87\% \\
\addlinespace[0.5em]
47 & $\begin{bmatrix}
0 & 0 & 1 & 0 \\ 
1 & 1 & 1 & 1 
\end{bmatrix}$ & 2.04\% && 
60 & $\begin{bmatrix}
0 & 0 & 1 & 1 \\ 
1 & 1 & 0 & 0 
\end{bmatrix}$ & 1.86\% \\
\addlinespace[0.5em]
\bottomrule
\end{tabular}
\end{table}

With regard to device ordering, we find that group 135 (with a starting smartphone session) is found more as often than the bitwise complement group 30 (with an ending smartphone session). We find similar relationships for other groups and in general we find that, in terms of all multidevice sessions, smartphone sessions tend more often to precede tablet sessions than vice versa. Google \cite{google2012} also identifies smartphones as the most common starting point for multidevice usage (when considering all sequential mutltidevice sessions).

In terms of diversity, we find that at least 25 groups (10\% of all possible groups) have frequencies of 1\% or greater and 138 groups (54\% of all groups) have at least one session thus suggesting significant diversity in the configuration of multidevice usage sessions. This significant diversity again reinforces the implication that app and interaction designer should consider a large variety of complex multidevice interactions and usage in future designs.

To account for differences in the number of multidevice sessions per user, we also calculate the relative frequency of each group for each user and then average these percentages. Table \ref{md_prototype_groups} also details the ten most frequent groups by this mean frequency per user. Nine of the top ten groups are the same as in overall analysis.  However, we find much smaller differences in device ordering than in the overall analysis. For example, groups 135 and 30 have almost equal mean frequencies of 3.62\% and 3.37\% respectively. Thus suggesting no preference in device ordering on a user level.

Finally we examine the types of apps used within the most frequent group (group 15) and the second most frequent group (group 240) compared to all other groups. Specifically we calculate the differences in normalized usage times by app category between group 15 and 240 (hereafter G15 and G240) sessions and the complements of group 15 and group 240 (hereafter C15 and C240) sessions for both smartphones and tablets. 

For G15, as expected, we find the sparse usage of smartphones to be related to app categories with frequent notifications such as news or social networking (83\% and 41\% more usage in G15) rather than app categories such as video or entertainment (280\% and 87\% less usage). Interestingly tough, we find that the usage of tablets in G15 to be related more with gaming (33\% more usage) rather than video (77\% less usage). Thus game developers should be especially aware that tablet users might often need to interrupt games for frequent smartphone notifications.

For G240, we find the sparse usage of tablets to be related to app categories such as productivity (email and web browsing) (60\% more usage in G240) rather than app categories such as games or video (160\% and 580\% less usage). Whereas, we find that the usage of smartphones in G240 to be related to categories such as photos and gallery, maps and navigation, or video (85\%, 70\%, and 62\% more usage) rather than app categories such as camera, gaming, or social networking (252\%, 76\%, and 36\% less usage).

Thus we find that although G15 and G240 can be illustrated as inverses, the types of apps commonly used in these sessions are different. In other words, the app categories that are commonly used on smartphones in G15 are not necessarily commonly used on Tablets in G240 and vice versa. Therefore, understanding the dominant device of a multidevice session is likely to be useful in, for example, inferring the types of apps used in that session. This type of inference is useful in cases where app sessions themselves might be inferred from other sources and thus no direct app type information is available.

\subsection{Smartphone Usage with and without Simultaneous Tablet Usage} \label{single_multi_device_simul}
Beyond the interplay within single multidevice usage sessions, we study the diversity of a user's smartphone usage behavior across different types of usage sessions. Specifically we study whether multidevice users use their smartphones differently when they are also using a tablet device in the same usage session\footnote{We note that multidevice users' tablet usage behavior in different situations is also interesting but not as easily studied because certain session types, for example tablet pure usage sessions, are relatively rare in practice. Thus we omit such analysis.}. We can formalize this question as comparing smartphone usage behavior between smartphone pure usage sessions vs. smartphone mixed usage sessions. 

In practice, our comparison consists of testing for statistically significant differences in normalized usage times\footnote{The times must be normalized because the total usage times for these two sessions types are not equal.} of both app categories and individual apps between the session types. In addition, since differences in app usage times are affected by the time of day and different session types are more prevalent during certain times, we limit the analysis to the typical evening period between 17:00 and 24:00.\footnote{We also performed the analysis without this constraint and found similar results.}

The examined app categories are custom categories that unify the category differences between application stores of different platforms (iOS and Android). In the analysis we only test app categories that at least 50\% of users used. This threshold keeps the sample sizes from becoming too small, thus rendering the tests unreliable. These custom categories are listed in Table \ref{app_categories_paired}. The individual apps are a collection of popular apps (again that at least 50\% of users used). The apps are listed in Table \ref{individual_apps_paired}.  Refer to Table \ref{tab_a4} in Appendix A for general usage statistics for the app categories and individual apps.

To test for significant differences, we use a percentile bootstrap method for paired data that compares difference scores\footnote{This difference score approach, as opposed to a marginal distributions approach, tells about the typical differences for a randomly selected user. Refer to section 5.9.9 in \cite{wilcox2012}} of the 20\% trimmed mean (section 5.9.12 in \cite{wilcox2012}). The percentile bootstrap method is selected because the method is robust under non-normality and heterogeneity of variances. Furthermore, we use the 20\% trimmed mean as the location measure of the tests because the distributions have significant outliers. Thus the null hypothesis is that the 20\% trimmed mean of the distribution of the difference scores of the groups is equal to zero. All tests are performed in R with the bootdpci command from \cite{wilcox2012}. Furthermore, for significant results ($p < 0.05$) we report a robust effect size measure, $\xi$, based on a generalization of explanatory power \cite{wilcox2011}. We interpret effect sizes $\xi > 0.15, 0.35, 0.50$ as small, medium and large effects respectively, as suggested in section 5.3.4 in \cite{wilcox2012} and we report the absolute value of effect sizes. All effect sizes were found in R with the yuenv2 command from \cite{wilcox2012}. 

In terms of application categories, column C1 in Table \ref{app_categories_paired} details the tests results. We find significant differences in the normalized usage time of several app categories. In all cases we find higher usage time in the smartphone pure sessions. In the cases of Camera, Photo and Gallery, and Video the reduced usage in mixed sessions can be explained by users preferring the larger tablet screen, which is typically optimized for viewing high resolution images. All three categories also have large effect sizes. In terms of individual app usage, column C2 in Table \ref{individual_apps_paired} details the tests results. We find significant differences in the usage time of Web Browsers, Email, and Messengers. Again, we find higher usage among the smartphone pure sessions.

Overall, the results suggest that users do use their smartphone differently when they are also using a tablet device in the same usage session. However the usage differences vary significantly for both application categories and individual apps. These differences suggest that platform or app developers could optimize the user experience in cases where, for example, the smartphone can detect or infer whether a tablet is being used simultaneously. Such optimizations would be an logical extension of already existing contextual optimizations (such as detecting significant location like home or work) to the case of using multiple devices. For example, when viewing gallery photos on a smartphone with a nearby tablet device, the smartphone can suggest that the tablet be automatically woken up with the same photo in the gallery loaded so the larger screen tablet can be utilized for viewing.

\begin{table}[!t]
\centering
\begin{threeparttable}
\caption{Smartphone Usage across Session Types - App Category Tests Results}
\label{app_categories_paired}
\begin{tabular}{@{}ll@{}}
\toprule
Category Name & C1 P-values\tnote{a} \\
\midrule
Other & 0.776 PS\textgreater MS \\
Shopping & \textbf{0.036* PS\textgreater MS (0.66)} \\
Health & -\tnote{b} \\
Productivity & 0.122 PS\textgreater MS \\
Maps and Navigation & - \\
Entertainment and Sports & 0.670 PS\textgreater MS \\
Camera & \textbf{0.000*** PS\textgreater MS (0.71)} \\
Weather & \textbf{0.004** PS\textgreater MS (0.38)} \\
Photos and Gallery & \textbf{0.000*** PS\textgreater MS (0.81)} \\
E-Books & - \\
Games & 0.432 PS\textgreater MS \\
Finance & 0.152 PS\textgreater MS \\
Social Networking & 0.066 PS\textgreater MS \\
Music and Audio & - \\
Video & \textbf{0.026* PS\textgreater MS (0.66)} \\
News & - \\
\bottomrule
\end{tabular}
    \begin{tablenotes}
      \small
      \item[a] For significant results the significance level, relationship between trimmed means, and robust effect size are also reported.
        \item[b] Indicates that not enough users of the category to reach threshold of at least 50\% usage. 
    \end{tablenotes}
\end{threeparttable}
\end{table}

\begin{table}[!t]
\centering
\begin{threeparttable}
\caption{Smartphone Usage across Session Types - Individual App Tests Results}
\label{individual_apps_paired}
\begin{tabular}{@{}ll@{}}
\toprule
App Name & C2 P-values\tnote{a} \\
\midrule
Other & \textbf{0.000*** MS\textgreater PS (0.16)} \\
Web Browsers & \textbf{0.040* PS\textgreater MS (0.19)} \\
YouTube & -\tnote{b} \\
Email Apps & \textbf{0.002** PS\textgreater MS (0.41)} \\
Facebook & 0.088 PS\textgreater MS \\
Messenger Apps & \textbf{0.008** PS\textgreater MS (0.12)} \\
\bottomrule
\end{tabular}
    \begin{tablenotes}
      \small
      \item[a] For significant results the significance level, relationship between trimmed means, and robust effect size are also reported.
      \item[b] Indicates that not enough users of the app to reach threshold of at least 50\% usage. 
    \end{tablenotes}
\end{threeparttable}
\end{table}

\subsection{Smartphone and Tablet Usage Substitution}
In this section we study the potential substitution effects between smartphones and tablets. More specifically, we study whether tablet usage is primarily usage shifted from smartphones (thus substituting the tablet for the smartphone), truly novel usage, or a combination of both. Furthermore, we study which specific activities are part of any shifted usage or novel usage.

Undoubtedly, the most sound approach to study such substitution effects would be to compare smartphone usage before and after the acquisition of a new tablet. However given a one month panel, relatively few users acquire a tablet during the observation period. Thus, we instead use an indirect approach to measuring these effects. Specifically, we compare the usage behavior of multidevice users (dataset 1, hereafter MD users) and non-multidevice users (dataset 2, hereafter NMD users). Thus under the assumption that these users are relatively similar, we can identify the tablet usage of MD users as purely novel usage, usage where a tablet was substituted for a smartphone, or a combination of both. We perform this identification by studying two distinct issues. First we study whether MD users have significantly more overall usage (across all devices) compared to NMD users (thus indicating novel usage).  Second, we study whether MD users have significantly less smartphone usage than NMD users (thus indicating substitution). Furthermore, we make these comparisons first with regard to all usage, and then for individual app categories and individual apps. We use the same threshold for including app categories and individual apps as in Section \ref{single_multi_device_simul}.

In practical terms, performing both of these comparisons is again reducible to testing for significant differences between usage distributions. To perform these tests, we use a percentile bootstrap method that compares the 20\% trimmed means (section 5.4 in \cite{wilcox2012}). Again we use the 20\% trimmed mean as the location measure because of significant outliers. Thus the null hypothesis is equality of the 20\% trimmed means between the groups. All tests are performed in R with the trimpb2 command from \cite{wilcox2012}. And again, for significant results ($p < 0.05$) we also report a robust effect size measure.

First, we compare the overall usage distributions (all usage across all devices) of the groups, Column C3 in Table \ref{total_app_time} details the test result. We find that group MD has significantly higher trimmed mean usage time, 244.64 compared to 172.81 minutes per day. Second, we compare the smartphone usage distributions of the groups (all usage from smartphones), column C4 in Table \ref{total_app_time} details the test result. We find group NMD has significantly higher trimmed mean usage time, 172.81 compared to 138.00 minutes per day. Thus these results suggest an overall combination of substitution and novel usage. Specifically, the results suggest that of the mean 71.83 minutes per day of tablet usage about 48\% is substitution and 52\% is novel usage. Finally, in terms of magnitude, the robust effect sizes suggest weak to moderate effects. This combination of substitution and novel usage have been previously observed in the case of smartphone and PC users where usage information was collected through semi-structured interviews \cite{matthews2009}.

\begin{table}[!t]
\centering
\begin{threeparttable}
\caption{Smartphone and Tablet Substitution - Total App Time Tests Results}
\label{total_app_time}
\begin{tabular}{@{}lll@{}}
\toprule
 & C3 P-values\tnote{a} & C4 P-values\tnote{a}\\
\midrule
Total App Time & \textbf{0.001*** MD\textgreater NMD (0.33)} & \textbf{0.015* NMD\textgreater MD (0.22)} \\
\bottomrule
\end{tabular}
    \begin{tablenotes}
      \small
      \item[a] For significant results the significance level, relationship between trimmed means, and robust effect size are also reported.
    \end{tablenotes}
\end{threeparttable}
\end{table}

Next, we perform the same comparisons but with regard to app usage within app categories rather than overall usage. Column C5 in Table \ref{app_categories_unpaired} details the tests results for app categories across all devices. Whereas column C6 in Table \ref{app_categories_unpaired} details the tests results for app categories from smartphones. 

\begin{table}[!t]
\centering
\begin{threeparttable}
\caption{Smartphone and Tablet Substitution - App Category Tests Results}
\label{app_categories_unpaired}
\begin{tabular}{@{}lll@{}}
\toprule
Category Name & C5 P-values\tnote{a} & C6 P-values\tnote{a}\\ 
\midrule
Other & \textbf{0.013* MD\textgreater NMD (0.26)} & 0.443 NMD\textgreater MD \\
Shopping & 0.490 MD\textgreater NMD & 0.337 NMD\textgreater MD\\
Health & 0.054 MD\textgreater NMD & 0.050 MD\textgreater NMD\\
Productivity & \textbf{0.016* MD\textgreater NMD (0.20)} & 0.083 NMD\textgreater MD\\
Maps and Navigation & 0.449 MD\textgreater NMD & 0.835 NMD\textgreater MD\\
Entertainment and Sports & 0.115 MD\textgreater NMD & 0.202 NMD\textgreater MD\\
Camera & 0.190 MD\textgreater NMD & 0.331 MD\textgreater NMD\\
Weather & 0.348 MD\textgreater NMD & 0.634 MD\textgreater NMD\\
Photos and Gallery & 0.164 MD\textgreater NMD & 0.315 MD\textgreater NMD\\
E-Books & 0.061 MD\textgreater NMD & 0.520 NMD\textgreater MD\\
Games & 0.216 MD\textgreater NMD & \textbf{0.004*** NMD\textgreater MD (0.29)} \\
Finance & 0.103 MD\textgreater NMD & 0.880 MD\textgreater NMD\\
Social Networking & 0.334 MD\textgreater NMD & 0.850 MD\textgreater NMD\\
Music and Audio & 0.359 NMD\textgreater MD & 0.057 NMD\textgreater MD\\
Video & \textbf{0.011* MD\textgreater NMD (0.28)} & 0.201 NMD\textgreater MD\\
News & 0.173 MD\textgreater NMD & 0.877 NMD\textgreater MD\\
\bottomrule
\end{tabular}
    \begin{tablenotes}
      \small
      \item[a] For significant results the significance level, relationship between trimmed means, and robust effect size are also reported.
    \end{tablenotes}
\end{threeparttable}
\end{table}

Interestingly, we find relatively few categories with significant novel usage or substitution, only the productivity and video categories indicate potential novel usage while only the games category indicates substitution. The substitution of a tablet for smartphone while playing games is unsurprising given that many games require touch based gestures and interaction that is more easily performed on the larger tablet display. Interestingly though, the productivity and video categories together only account for about 38\% of the total novel usage and games only accounts for about 20\% of total substituted usage. Thus other categories are likely also affected though not to a statistically significant level. Also we note that in all results from this comparison the effect sizes are relatively small.

Finally, we perform the comparisons with regard to app usage of a collection of individual apps. Column C7 in Table \ref{individual_apps_unpaired} details the tests results for individual apps across all devices and column C8 in Table \ref{individual_apps_unpaired} details the tests results for individual apps from smartphones. The results indicate web browsers and email apps as novel usage which is expected given that they are both classified as productivity apps. However, the most surprising result is that the YouTube app is not significant despite the significance of the larger video category. This might be in part due to the typically shorter and lower resolution clips on YouTube that do not benefit as much from a larger tablet display, as compared to other applications in the video category such as Netflix, Flixter, and Google Play Movies.

\begin{table}[!t]
\centering
\begin{threeparttable}
\caption{Smartphone and Tablet Substitution - Individual App Tests Results}
\label{individual_apps_unpaired}
\begin{tabular}{@{}lll@{}}
\toprule
App Name & C7 P-values\tnote{a} & C8 P-values\tnote{a}\\
\midrule
Other & \textbf{0.005** MD\textgreater NMD (0.27)} & \textbf{0.002** NMD\textgreater MD (0.27)} \\
Web Browsers & \textbf{0.003** MD\textgreater NMD (0.25)} & 0.057 NMD\textgreater MD \\
YouTube & 0.140 MD\textgreater NMD & 0.148 NMD\textgreater MD \\
Email Apps & \textbf{0.029* MD\textgreater NMD (0.19)} & 0.781 MD\textgreater NMD \\
Facebook & 0.133 MD\textgreater NMD & 0.772 NMD\textgreater MD \\
Messenger Apps & 0.969 NMD\textgreater MD & 0.913 NMD\textgreater MD \\
\bottomrule
\end{tabular}
    \begin{tablenotes}
      \small
      \item[a] For significant results the significance level, relationship between trimmed means, and robust effect size are also reported.
    \end{tablenotes}
\end{threeparttable}
\end{table}

Importantly, our analysis of substitution effects relies on the assumption that the users of the two groups are relatively similar. This assumption is certainly not strictly true since certain types of users (basic, advanced, etc.) are both more likely to acquire a tablet device and to, for example, use certain applications more often. However, the differences in user types are unlikely to be large enough to cause all of the observed effects. We find that only household income is statistically significantly different ($p < 0.05$) in the demographics and smartphone technographics described in Table \ref{panel_demos} between the groups. Furthermore, we look to address these concerns in future work (refer to Section \ref{conclusions}).

\section{Discussion}\label{discussion}
We briefly reiterate a few implications of the analysis results for two different mobile ecosystem players and we discuss our views on replicability and generalizability.

For developers and designers of mobile apps, our results have implications in terms of illustrating both the ubiquity of and diversity within multidevice mobile interaction. For example, our results indicate that, for our panelists, over 50\% of mobile device usage time can be considered multidevice usage. This ubiquity implies that developers and designers should {\it already} be thinking and designing in terms of multidevice mobile sessions. Towards this end, developers and designers can formalize their thinking through the application of already available multidevice interaction frameworks such as the 4C framework \cite{sorensen2014}.

Whereas the diversity within multidevice interaction suggests that multidevice solutions should adapt to a particular user's usage patterns (as less adaptive solutions would not be able to handle this diversity). Relatedly this diversity will likely grow as the number of device types increases and new device types continue to fill small under-served niches (the smartwatch \cite{pizza2016} being a clear example).

Additionally for mobile developers and designers, we also provide potential targets for improved mobile interaction design such as certain smartphone usage sessions where simultaneous usage of a tablet device can be inferred (as we illustrated based on smartphone app usage). For example, a user might start watching a video on their smartphone and the smartphone can suggest that the video be played directly on the tablet if simultaneous tablet usage is inferred. Such contextual inference can be viewed as helping users exploit the strengths and weaknesses of different device types without the need for the user to consciously consider, for example, device switches (as is the current case with systems such as Apple airplay).

Finally, in relation to related work, given our device based data approach, our work can be viewed as complimentary to expert survey based studies of design and development challenges for multidevice interaction such as \cite{dong2016} and \cite{grubert2016}. For example, \cite{dong2016} specifically identified ``unverified assumptions about multidevice use'' as a challenge in multidevice experience design. They detail that one expert believed that users seldom access multiple mobile platforms (for example Android smartphone and iOS tablet). However we found that 14\% of panelists had different platforms on their smartphones compared to their tablets. Though as we mention later, we have to be cautious with generalizations.

Whereas for advertisers, our results have implications in terms of understanding in-app advertising in the smartphone and tablet multidevice context. For example advertisers can understand the relative ad inventories between specific device types of users. Additionally we show that, given fixed ad refresh rates (in short timescales) and the effect of impression time on ad recognition and recall \cite{goldstein2011}, the differences in smartphone and tablet session characteristics (length and frequency) suggest simple metrics such as number of ad impressions might need to be complemented with more complex metrics such as multidevice ad impression time.

Finally, we also discuss our views on replicability and generalizability. Replicability and generalizability are important concerns in panel based studies in general. Furthermore, given the rapidly changing nature of mobile devices, these concerns are especially relevant for mobile device based panels and we attempt to alleviate some of these concerns by having a relatively diverse multi-platform panel. However, overall we agree with Church et al. \cite{church2015} that these studies are mainly about the panel populations themselves and the value is primarily in allowing researchers and the community to compare and contrast different experiences with different user populations to build a comprehensive understanding of the variety of mobile users and their unique behaviors. In this sense our multidevice usage study provides an initial benchmark that we hope spurs further investigations.

\section{Conclusions and Future Work}\label{conclusions}
In this paper, we provided a first look at multidevice usage by examining the smartphone and tablet usage of a US-based user panel from February 2015. Our contributions include both a descriptive analysis of the panel's multidevice usage and novel methods for examining multidevice usage in general. 

We present a basic method for aggregating app sessions into multidevice usage sessions through well known Allen's interval relations. 

Then we provide extensive descriptive statistics of both the multidevice and single device usage sessions. The statistics indicate, for example, that about 50\% of total device usage time is multidevice usage, 40\% smartphone only usage, and 10\% tablet only usage.  

We describe and apply a novel method to study diversity in temporal patterns in multidevice sessions. The results indicate that about 35\% of multidevice sessions are dominated by a single device with only sparse usage of the second device. However, we also find significant diversity with 25 different temporal patterns having at least 1\% share of multidevice sessions.

We compare the smartphone usage of users inside and outside multidevice sessions. Paired statistical tests indicate that in general users do use their smartphones differently during multidevice sessions compared to single device sessions. However the magnitude of the differences depends on the specific applications used.

Finally, we study usage substitution between smartphone and tablet devices by comparing the usage of the multidevice panel and a supplementary panel of users without tablet devices. The results indicate that tablet usage is likely a combination of usage shifted from a smartphone (thus substituting the tablet for the smartphone) and truly novel usage.  Quantitatively, the results suggest that tablet usage is about equally divided between these cases, 48\% and 52\% respectively. Furthermore, we find that productivity apps such as web browsing and email and video apps account for a significant portion of novel usage, while game apps account for a significant portion of shifted usage.

In terms of general future research, a natural direction would be to expand the multidevice analysis to include personal computer usage. Personal computer usage data was collected from panel users but considered out of scope for this paper. As mentioned in Section \ref{datasets}, a difficulty in such an expansion is that the simple foreground app session definition is not directly transferable to the personal computer context. Though we plan to examine this difficulty in future work.

For future research in device substitution effects, the demographic information might allow a propensity score matching \cite{rosenbaum1983} approach to user group comparison. Such an approach would strengthen resulting conclusions. Furthermore since the panels are continuously ongoing, eventually the gold standard comparison of usage before and after device acquisition might be feasible.
\section*{Acknowledgment}

The authors would like to thank Verto Analytics for providing the data, and Heikki H{\"a}mm{\"a}inen and Kalevi Kilkki for providing feedback during the process.


\bibliography{main}

\newpage
\appendix
\section{} \label{App:AppendixA}

\begin{table}[!htbp]
\scriptsize
\centering
\begin{threeparttable}
\caption[Shares of app category and app usage time]{Shares of app category and app usage time\tnote{a}}
\label{tab_a4}
\begin{tabular}{@{}lrrr@{}}
\toprule
 & \multicolumn{2}{c}{{\bf Dataset 1}} & \multicolumn{1}{c}{{\bf Dataset 2}} \\ \midrule
{\bf Category Name} & {\bf Smartphone} & {\bf Tablet} & {\bf Smartphone} \\
Other & 37.97 \% & 42.84 \% & 42.70 \% \\
Social Networking & 31.62 \% & 5.77 \% & 20.61 \% \\
Games & 8.09 \% & 27.81 \% & 13.22 \% \\
Video & 6.06 \% & 12.27 \% & 3.93 \% \\
Productivity & 4.21 \% & 2.09 \% & 3.20 \% \\
Shopping & 3.72 \% & 1.30 \% & 3.57 \% \\
Photos and Gallery & 1.55 \% & 0.18 \% & 0.70 \% \\
E-Books & 1.47 \% & 0.62 \% & 1.81 \% \\
Entertainment and sports & 1.25 \% & 3.03 \% & 6.07 \% \\
Music and Audio & 1.08 \% & 0.40 \% & 1.48 \% \\
Maps and Navigation & 0.67 \% & 0.11 \% & 0.60 \% \\
Camera & 0.55 \% & 0.08 \% & 0.41 \% \\
Health & 0.54 \% & 0.03 \% & 0.40 \% \\
News & 0.48 \% & 3.16 \% & 0.49 \% \\
Finance & 0.38 \% & 0.16 \% & 0.47 \% \\
Weather & 0.36 \% & 0.15 \% & 0.33 \% \\
{\bf App Name} &  &  &  \\
Other & 49.69 \% & 62.04 \% & 59.63 \% \\
Facebook & 16.43 \% & 4.60 \% & 10.57 \% \\
Messenger Apps & 15.19 \% & 1.17 \% & 10.04 \% \\
Web Browsers & 12.59 \% & 18.33 \% & 13.51 \% \\
Email Apps & 4.16 \% & 2.03 \% & 3.19 \% \\
YouTube & 1.94 \% & 11.83 \% & 3.06 \% \\ \bottomrule
\end{tabular}
    \begin{tablenotes}
      \small
      \item[a] The values are calculated over the whole datasets
    \end{tablenotes}
\end{threeparttable}
\end{table}

\begin{table}[!htbp]
\scriptsize
\centering
\begin{threeparttable}
\caption[Usage session construction statistics]{Usage session construction statistics for dataset 2}
\label{tab_a1}
\begin{tabular}{@{}lr@{}}
\toprule
 & {\bf Smartphone usage} \\ \midrule
{\bf Number of app sessions} & 2917208 \\
{\bf Number of single device usage sessions} & 629471 \\
{\bf Share of Allen's relations} &  \\
precedesWithinTW & 8.23 \% \\
meets & 41.95 \%  \\
metBy & 41.60 \%  \\
precededByWithinTW & 8.23 \% \\ \bottomrule
\end{tabular}
\end{threeparttable}
\end{table}

\begin{table}[!htbp]
\scriptsize
\centering
\begin{threeparttable}
\caption[Basic statistics on the whole dataset level for dataset 2]{Basic statistics on the whole dataset level for dataset 2}
\label{tab_a2}
\begin{tabular}{@{}lrr@{}}
\toprule
 & {\bf Usage session length (s)} & {\bf App sessions per usage session} \\ \midrule
mean & 382 & 4.61 \\
median & 62 & 2.00 \\
std. dev. & 2156 & 8.44 \\
\bottomrule
\end{tabular}
\end{threeparttable}
\end{table}

\begin{table}[!htbp]
\scriptsize
\centering
\begin{threeparttable}
\caption[Basic statistics on per user level for dataset 2]{Basic statistics on per user level for dataset 2\tnote{a}}
\label{tab_a3}
\begin{tabular}{@{}lrrrr@{}}
\toprule
\multicolumn{1}{r}{} & {\bf \begin{tabular}[c]{@{}r@{}}Usage session\\ length (s)\end{tabular}} & {\bf \begin{tabular}[c]{@{}r@{}}App sessions per\\ usage session\end{tabular}} & {\bf \begin{tabular}[c]{@{}r@{}}Usage sessions\\ per day\end{tabular}} & {\bf \begin{tabular}[c]{@{}r@{}}Interaction time\\ per day (min)\end{tabular}} \\ \midrule
mean & 171 & 2.57 & 47.12  & 298.87 \\
median & 66 & 3.00 & 42.06 & 261.09 \\
std. dev. & 839 & 1.33 & 27.76 & 205.64 \\
\bottomrule
\end{tabular}
    \begin{tablenotes}
      \small
      \item[a] Usage session length and app sessions per usage session statistics are calculated from the per user medians of these metrics.
    \end{tablenotes}
\end{threeparttable}
\end{table}

\end{document}